\pdfoutput=1

\documentclass[journal]{IEEEtran}
\ifCLASSINFOpdf
   \usepackage[pdftex]{graphicx}
   \graphicspath{{../pdf/}{../jpeg/}}
   \DeclareGraphicsExtensions{.pdf,.jpeg,.png}
\else
\fi
\usepackage{latexsym}
\usepackage[cmex10]{amsmath}
\newcommand{\str}[1]{#1}	

\usepackage{multirow}
\usepackage{subfigure}

\newcommand{\MYfooter}{\smash{\scriptsize
\hfil\parbox[t][\height][t]{\textwidth}{\centering
~\\
~}\hfil\hbox{}}}

\newcommand{\MYarxivheader}{\smash{\scriptsize
\hfil\parbox[t][\height][t]{\textwidth}{\centering
(c) 2015 IEEE. Personal use of this material is permitted. Permission from IEEE must be obtained for all other users, including reprinting/ republishing this material for advertising or promotional purposes, creating new collective works for resale or redistribution to servers or lists, or reuse of any copyrighted components of this work in other works.}\hfil\hbox{}}}

\makeatletter

\def\ps@headings{%
\def\@oddhead{\mbox{}\scriptsize\rightmark \hfil \thepage}
\def\@evenhead{\scriptsize\thepage \hfil \leftmark\mbox{}}
\def\@oddfoot{\MYfooter}%
\def\@evenfoot{\MYfooter}}

\def\ps@IEEEtitlepagestyle{%
\def\@oddhead{\MYarxivheader}%
\def\@evenhead{\scriptsize\thepage \hfil \leftmark\mbox{}}%
\def\@oddfoot{\MYfooter}%
\def\@evenfoot{\MYfooter}}

\makeatother

\begin{document}
%
\title{A Batteryless Sensor ASIC for Implantable Bio-impedance Applications}
%
%
%

\author{Saul~Rodriguez,~\IEEEmembership{Member,~IEEE,}
      Stig~Ollmar,~\IEEEmembership{Senior Member,~IEEE,}
      Muhammad~Waqar,~\IEEEmembership{Student Member,~IEEE,}
       and~Ana~Rusu,~\IEEEmembership{Member,~IEEE}

\thanks{Manuscript received January 14, 2015. Support from VR through the ErBio and Mi-SoC projects is gratefully acknowledged.}

\thanks{S. Rodriguez, M. Waqar, and A. Rusu are with the KTH Royal Institute of Technology, School of ICT, Kista, Sweden (email: saul@kth.se;  mwhus@kth,se; arusu@kth.se)}
\thanks{S. Ollmar is with the Karolinska Institute (email: stig.ollmar@ki.se)}}
%
%

\markboth{Journal of \LaTeX\ Class Files,~Vol.~11, No.~4, December~2012}%
{Shell \MakeLowercase{\textit{et al.}}: Bare Demo of IEEEtran.cls for Journals}
%



\maketitle

\begin{abstract}
The measurement of the biological tissue's electrical impedance is an active research field that has attracted a lot of attention during the last decades. Bio-impedances are closely related to a large variety of physiological conditions; therefore, they are useful for diagnosis and monitoring in many medical applications. Measuring living tissues, however, is a challenging task that poses countless technical and practical problems, in particular if the tissues need to be measured under the skin. This paper presents a bio-impedance sensor ASIC targeting a battery-free, miniature size, implantable device, which performs accurate 4-point complex impedance extraction in the frequency range from 2~kHz to 2~MHz. The ASIC is fabricated in 150~nm CMOS, has a size of 1.22~mm~$\times$~1.22~mm and consumes 165~$\mu$A from a 1.8~V power supply. The ASIC is embedded in a prototype which communicates with, and is powered by an external reader device through inductive coupling. The prototype is validated by measuring the impedances of different combinations of discrete components, measuring the electrochemical impedance of physiological solution, and  performing  \textit{ex~vivo} measurements on animal organs. The proposed ASIC is able to extract complex impedances with around 1~$\Omega$\ resolution; therefore enabling accurate wireless tissue measurements.
\end{abstract}


%
\IEEEpeerreviewmaketitle

\section{Introduction}
%
%
%
%
\IEEEPARstart{T}{he} electrical impedance of the biological tissue, commonly referred as bio-impedance, is a function of physiological processes and therefore offers interesting opportunities for monitoring a variety of bio-markers. Its potential use in medical applications was already identified more than a century ago \cite{Grimnes2015}. Currently, bio-impedance research is still a growing field of study and its advances find spread use in many medical applications.  Among others, bio-impedance characterization is the principle behind Electrical Impedance Tomography (EIT) \cite{Holder2005a}, which has wide use in applications such as lung function monitoring, breast cancer detection, cervical intraepithelial neoplasia prediction,  brain imaging, etc. Bio-impedance is also being successfully used to detect different types of skin cancer \cite{Aberg2004a}. In addition, bio-impedance measurements are being used as part of complementary tools to remove artifacts in other kind of measurements such as in ECG \cite{Kim2014},\cite{VanHelleputte2012},\cite{Yazizioglu2010},\cite{VanHelleputte2015}. Typically, bio-impedance is extracted at a single frequency in these applications. Furthermore, surgical instrumentation controlled by bio-impedance measurements has been recently proposed \cite{Brendle2014}. Other applications include body fluid analysis and cell characterization for wearable and implantable health care applications \cite{Kubendran2014},\cite{Beckmann2009}, \cite{Lee2015}, Kelvin impedance sensing to detect biomolecular interactions \cite{Crescentini2014}, and myocardial ischemia detection \cite{Yufera2005}.

\begin{figure}[t!]
\centering
\includegraphics[width=8.5cm]{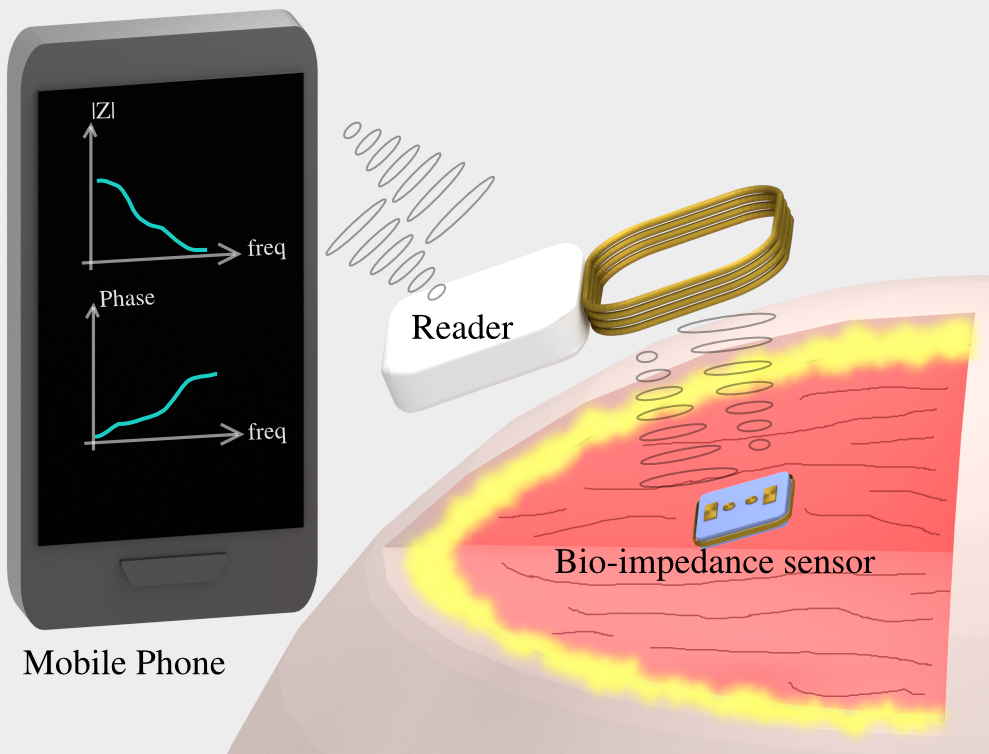}
\caption{Bio-impedance measuring system concept}
\label{FIG:INTRO}
\end{figure}
The measurement of bio-impedance using non-invasive methods, such as measurements on the skin, are plagued with artifacts \cite{Ollmar2013}  that make enormously difficult  to correctly characterize the tissue under test.  { Furthermore, a rigorous electrical characterization of internal organs' tissues under different physiological conditions needs the analysis of complex impedances at many frequency points over several decades (impedance spectroscopy).} Accordingly, implantable devices for measuring bio-impedance of tissues under the skin (i.e. muscle, bone, blood, etc.) {which are able to perform impedance spectroscopy} are necessary. Implantable sensors, however, have stringent limitations \cite{Ko2012} in size, power, and communication means which define to a great extent the way in which the bio-impedance sensor is built.


For practical and medical reasons, the size of an implantable device should be as small as possible, so miniaturization of the system becomes a consideration of paramount importance.  In addition, it should not have wires to the external world; therefore, it requires some means of wireless communication  \cite{Lee2013a}. Likewise, providing electrical power to the implantable device is a major challenge. The use of batteries is strongly discouraged since they have a limited lifespan and also have a strong impact on the form factor. Energy harvesting techniques such as inductive coupling are preferred since power transfer and wireless communication can be achieved at the same time  \cite{Cao2012},\cite{Iker2013},\cite{Zargham2014}. The inductive harvesters, however, are able to extract only limited amounts of power at large distances (low efficiencies); therefore,  ultra-low power design techniques must be employed to build the implantable electronic circuits.

\begin{figure*}[t!]
\begin{center}
\includegraphics[width=\textwidth]{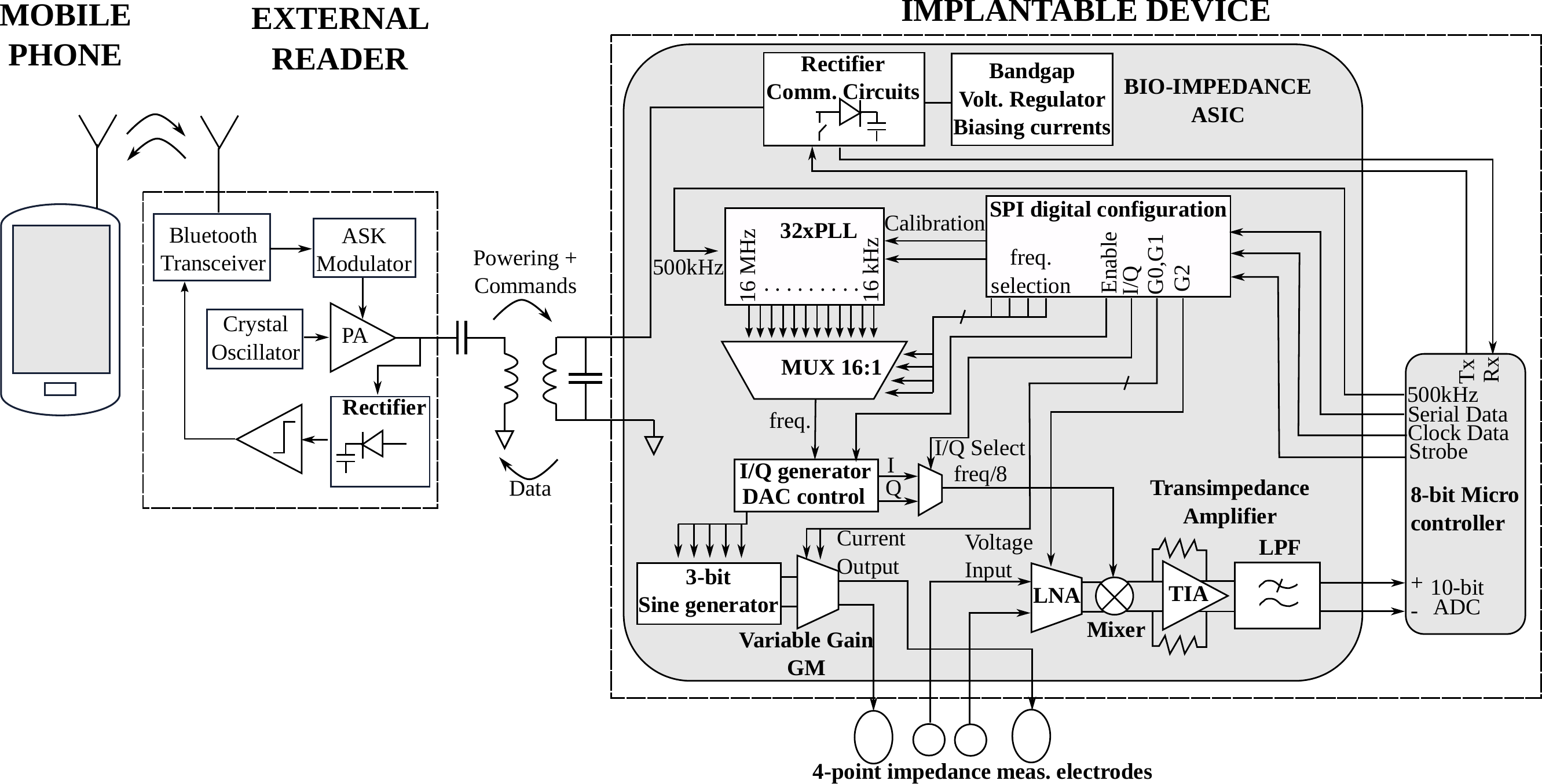}
\caption{Proposed bio-impedance sensing system}
\label{FIG:ARCHIT}
\end{center}
\end{figure*}

This paper presents a low power bio-impedance sensor ASIC that targets a miniature-size, battery-free implantable device. The ASIC is used in the bio-impedance sensing system presented in Fig.~\ref{FIG:INTRO}. 
{The objective of this system is to perform impedance spectroscopy of animal and human living tissues from 2~kHz to 2~MHz (11 points logarithmically spaced).}
The implantable device is externally powered by a reader device using inductive coupling. In addition, inductive coupling is used to establish two-way half-duplex communication  between the reader and the implantable device. The reader device is connected to an Android mobile phone by using Bluetooth. A software application including a graphic user interface (GUI) is used to send control orders to the implantable device, perform digital signal processing (post processing/calibration), and present measurement results.  The sensor ASIC is able to extract complex impedances from 2~kHz to 2~MHz with around 1~$\Omega$ {resolution} after calibration. 
{Table I shows a summary of the implantable device specifications.}
The paper is organized as follows. The complete bio-impedance sensing system architecture is described in Section II. The details of the different building blocks in the ASIC are shown in Section III. System integration and validation of the proposed bio-impedance sensor are presented in Section IV. Finally, conclusions are drawn in Section V.
\begin{table}[!t]
\renewcommand{\arraystretch}{1.5}
\caption{Design Specifications for the Implantable Sensor}
\centering
\begin{tabular}{c c}
\hline
\bfseries Name & \bfseries Specification\\
\hline\hline
Power source & Inductive coupling \\
Wireless Comm. & ASK over inductive coupling \\
Powering Distance & $>$ 4 cm  \\
Dimensions & Approx. 13~mm $\times$ 3~mm\\
Measured Quantity & Complex Impedance\\
Meas. Technique & 4-terminal  \\
Resolution & Approx. 1~$\Omega$\\
Freq. Range & 2~kHz - 2~MHz \\
\hline
\end{tabular}
\label{TABLE:SPEC}
\end{table}

\section{Implantable Sensor Architecture}

The bio-impedance sensor architecture depends completely on the impedance measurement technique that is used. Two-terminal auto-balance bridge with digital calibration \cite{Diao2012} is an amenable technique for integration in implantable devices since it requires only two electrodes. However, this technique has the drawback that it also measures the interface electrode impedances which can be much larger than the tissue's impedance. Removal of the electrode interface impedances is aggravated by mismatches which are difficult to estimate and therefore difficult to subtract. On the other hand, 4-terminal measurements alleviate the issue of the electrode interface impedance at the price of an extra pair of electrodes  \cite{Ackmann1993}. In this work, a 4-terminal impedance sensing method has been used to measure the tissue impedances. 

Fig. \ref{FIG:ARCHIT} shows all the components of the bio-impedance sensing system: Android phone, external reader, and implantable device. As can be seen in Fig.~\ref{FIG:ARCHIT}, the selected architecture is based on sinusoidal current injection, voltage sensing, and in-phase (I) and quadrature (Q) separation. The principle of operation is as follows.

A sinusoidal differential current $I$ with known amplitude, frequency, and phase is injected on the two external electrodes, and the differential  voltage drop $V$ is sensed on the other two internal electrodes. A complex impedance $Z=|Z|e^{j\theta}$ in the path of $I$ results in $V$ being proportional to $I$ and ``modulated''  both in amplitude and phase. Using phasor representation,  $V$ can be expressed as:
\begin{align}
V =|I|e^{j\omega t}\times Z = |I||Z|e^{j\omega t}e^{j\theta} = |I||Z|e^{j(\omega t +\theta)}
\label{EQ:V}
\end{align}
Extraction of the real and imaginary parts of the complex impedance is done by using in-phase (I) and quadrature (Q) demodulation, respectively. For the in-phase component, the sensed voltage is multiplied by $cos(\omega t)$ so that:
\begin{equation}
\begin{split}
 V_I  &=  |I||Z|e^{j(\omega t +\theta)}cos (\omega t) \\
  &= |I||Z|(cos (\omega t + \theta) + j sin( \omega t + \theta)) cos (\omega t) \\
  &= |I||Z| \Bigl( \frac{cos ( \theta)}{2} + \frac{cos (2\omega t + \theta)}{2} +  \\
	& \text{\hspace{5mm}}  j sin( \omega t + \theta)) cos (\omega t) \Bigl)
\end{split}
\label{EQ:VI}
\end{equation}
which after low-pass filtering becomes:
\begin{align}
V_{I,DC} = |I||Z|\frac{cos ( \theta)}{2}
\end{align}
Then, the real part can be recovered as:
\begin{align}
Re\{Z\} = |Z|cos ( \theta) = \frac{2  V_{I,DC}}{|I|}
\end{align}
For the quadrature component, the sensed voltage is multiplied by $sin (\omega t)$ instead, and after similar treatment, the imaginary part of $Z$ is:
\begin{align}
Im\{Z\} = |Z|sin ( \theta) = \frac{2  V_{Q,DC}}{|I|}
\end{align}

The I/Q demodulation requires the design of a very linear analog multiplier and generation of very clean I/Q sinusoidal signals. An alternative is the use of a current commutating passive mixer that is controlled by I/Q digital clock signals. The main advantages of this approach is that the passive mixer is inherently highly-linear, and does not consume any power. In addition, the generation of I/Q clocks are purely digital, and therefore easy to implement.

 The mixing process entails time-domain multiplication of $V$ by a square wave instead of a pure sinusoidal wave. In particular, the result of interest is the multiplication of $V$ by the fundamental harmonic $(4/\pi) cos(\omega t)$ for the in-phase component,  or by $(4/\pi) sin(\omega t)$ for the Q component. In addition, the demodulator chain amplifies the signal by a factor $G$ so that the quantization at the analog-to-digital converter (ADC) does not degrade the signal-to-noise ratio (SNR). When these details are taken in consideration, the expressions for $Re\{Z\}$ and $Im\{Z\}$ become:
\begin{align}
Re\{Z\} = |Z|cos ( \theta) =\frac{\pi}{2}  \frac{V_{I,DC}}{|I|G}
\end{align}
\begin{align}
Im\{Z\} = |Z|sin ( \theta) =\frac{\pi}{2}  \frac{V_{Q,DC}}{|I|G}
\end{align}

{Since the outputs of the I/Q demodulation are DC signals, it is unnecessary to demodulate the I/Q paths simultaneously by using two demodulator chains. Instead, it is possible to save area and power by using a single demodulator chain which is multiplexed in time. As shown in the block diagram of Fig.~\ref{FIG:ARCHIT}, this is easily implemented by using a digital configuration line which selects the I or Q clock that is used by the demodulator.}

The sinusoidal current source generation, frequency synthesis, and I/Q demodulation blocks are also integrated in the bio-impedance ASIC, as it can be seen in Fig.~\ref{FIG:ARCHIT}, and their design is described in detail in Section III. The following subsections provide brief descriptions of the external blocks of the system:  electrodes, circular coil, low power microcontroller, and external reader.

\subsection{Electrodes}

The geometry of the sensing electrodes is of paramount importance in impedance sensing. The limited dimensions of the implantable device demand the use of small geometries in which current injecting electrodes are close to the voltage sensing electrodes. Relatively small dimensions are advantageous in this application because electromagnetic fields are contained in a small space, and therefore, they allow only impedance extraction of the tissue of interest in the vicinity of the device. The geometry of the electrodes is also important because it defines the magnitude of the sensed impedance; hence, it sets specifications for the impedance range that the analog front-end circuits must be able to handle.

\begin{figure}[t!]
\centering
\includegraphics[width=5cm]{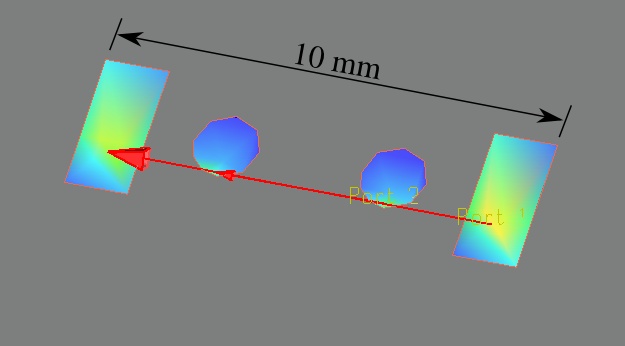}
\caption{EM simulation of the electrodes and tissue}
\label{FIG:ELECTRODE}
\end{figure}

\begin{figure}[t!]
\centering
\includegraphics[width=8cm]{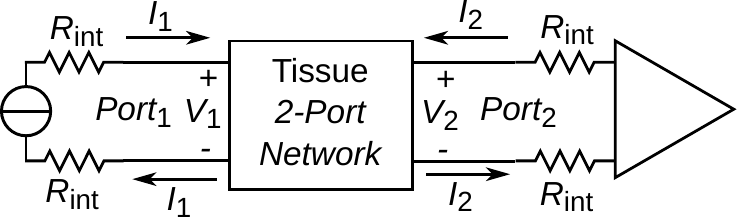}
\caption{Electrodes and tissue model as a 2-port network}
\label{FIG:PORT}
\end{figure}

In this work, the electrode configuration shown in Fig.  \ref{FIG:ELECTRODE} is used. The electrodes are made of gold (ENIG Ni/Au process) on a 0.75~mm hydrocarbon ceramic substrate Rogers 4000. An estimation of the tissue's impedance magnitude that can be detected by these electrodes is obtained by performing  Electromagnetic (EM) momentum simulation using Agilent ADS  in which the tissue is modeled as a 3 cm semi-conductor layer stacked on the top of the electrodes. Values of permittivity ($\varepsilon_r$) and conductivity ($\sigma$) are extracted from \cite{Damijan2006} and \cite{Rigaud1995} for different types of tissues. The electrodes are treated as a 2-port network in which the injecting current electrodes (rectangle shapes) are defined as differential port1, and the voltage sensing electrodes (circle shapes) are defined as differential port2 (Fig. \ref{FIG:PORT}). S-parameters are extracted and converted to Z-parameters. The parameter $Z_{21} = V_{2}/I_{1}$ which is calculated for $I_2 = 0$ corresponds to the sensed impedance. This simulation disregards the interface resistances $R_{int}$ between the electrodes and tissue which depends on many factors.  However, the results are valid since for these small geometries, the interface resistances only add in series to the total impedance seen from the injecting and sensing electrodes. More important, these resistances can be disregarded at the voltage sensing electrodes provided that the input amplifier presents very high impedance  (there are no voltage drops on the interface resistances since $I_2~\approx~0$). The lowest impedance $Z_{21}$, which takes a value of approximately 107 $\Omega$ at low frequencies, is obtained in the case of blood ($\sigma \approx 0.7$~S/m). Likewise, the largest impedance, which takes a simulated value of approximately 2491 $\Omega$ at low frequencies, is for the case of skeletal transversal muscle tissue ($\sigma \approx 0.03$~S/m).

\subsection{Circular coil}

Remote powering and communication of implantable devices by using inductive coupling has been successfully shown at distances from a few millimeters to several centimeters \cite{Catrysse2004}, \cite{Cheong2012}. Power levels in the order of several mW have been harvested with relatively good transfer efficiencies up to around 40\%. By using rotating magnets and low frequencies close to hundreds of Hz,  power quantities of several Watts can be harvested with transfer efficiencies as high as 50\% \cite{Jiang2013}. Nevertheless, so low frequencies demand large inductance values which occupy large sizes and therefore are not amenable for integration in miniature implantable devices. In addition, they make difficult the use of the inductive coupling for communication purposes since the communication bandwidth is commonly only a fraction of the operating frequency. In this work, inductive coupling is implemented by using standard circular coils and an operating frequency of 13.56~MHz (ISM band). The secondary coil in the implantable device is made of 0.25~mm diameter, enamelled copper wire. The circular coil  has  7 turns, and 12 mm diameter which results in an inductance of 1.3~$\mu$H.

\subsection{Low-power microcontroller}

A low-power microcontroller PIC16LF1823 from Microchip was selected for this work since it provides a robust and flexible digital solution in a small footprint. The microcontroller is housed in a QFN-32 package (4~mm~$\times$~4~mm), and it is used to control the bio-impedance sensor ASIC through a Serial Periphelal Interface (SPI) bus, to handle asynchronous serial communications with the reader, to provide a calibrated 500 kHz clock reference and 10-bit analog-to-digital conversion. The micro-controller consumes close to  2~$\mu$A when the firmware is running a waiting loop at 32-kHz, and around 100~$\mu$A when executing code at 500~kHz. 

\subsection{External Reader}

The external reader is a battery powered device that contains an RN42 Bluetooth transceiver module, a 13.56~MHz crystal oscillator, a Class-E power amplifier, the primary inductor composed of a 1.8~$\mu$H circular coil  (4~cm diameter), and an Amplitude Shift Keying (ASK) modulator/demodulator. The tasks of the reader are to provide wireless power to the implantable device through inductive coupling, and to relay data from/to  the Android phone and the implantable device.

\section{Bio-impedance ASIC Design}

The Bio-impedance ASIC was designed using a 150~nm 1.8~V CMOS process, and integrates the following blocks: SPI configuration, power management and communications, frequency synthesis, current signal generator, voltage sensing and I/Q demodulation, and low pass filter. The following subsections describe these building blocks.

\subsection{SPI Configuration}

The ASIC is configured through a SPI digital interface which is built by using a chain of registers. The inputs for this stage are a clear (reset) signal, serial data, clock data, and strobe. These configuration signals are provided by the microcontroller. A configuration word includes 2 calibration bits for the frequency synthesis block, 4 bits for frequency selection, 1 bit for enabling/disabling the signal source, 1 bit for selection of I/Q clock, and 3 bits that control the variable gain amplification.

\subsection{Power Management and Serial Data Transmission}

\begin{figure}[t!]
\centering
\includegraphics[width=8.5 cm]{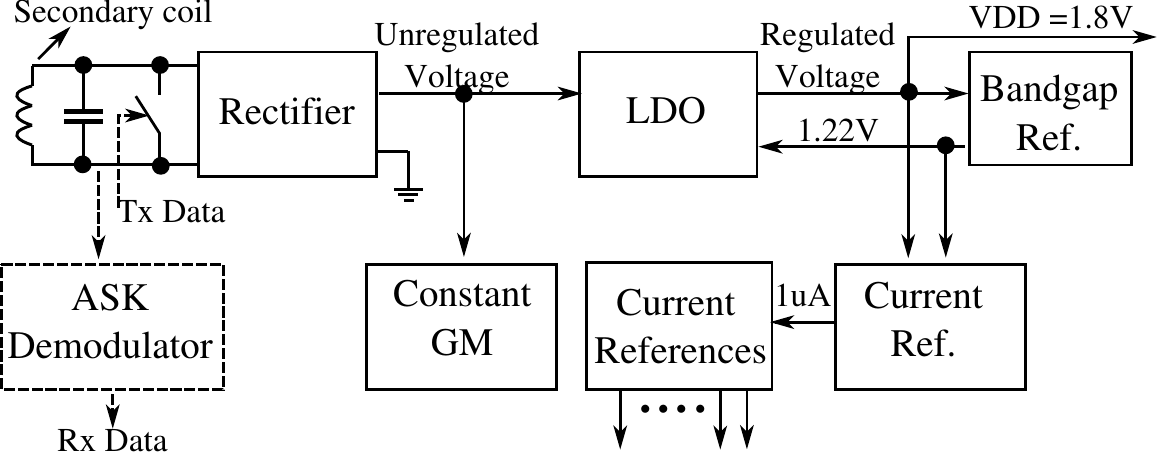}
\caption{Communications and power management block diagram}
\label{FIG:POWER}
\end{figure}

\begin{figure}[t!]
\centering
\includegraphics[width=6.5 cm]{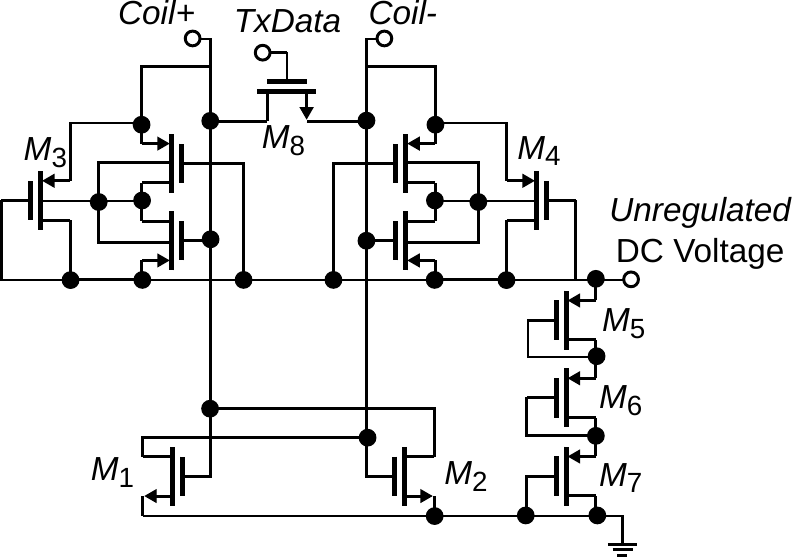}
\caption{Rectifier circuit}
\label{FIG:RECTIFIER}
\end{figure}

Fig. \ref{FIG:POWER} shows the power management and communication block diagram. The secondary circular coil and a parallel capacitor are tuned at the operating frequency 13.56~MHz. When the external reader is in the vicinity of the implantable device, AC voltage is induced in this LC tank circuit. A full-wave rectifier based on \cite{Ghovanloo2004} and shown in Fig. \ref{FIG:RECTIFIER} is used to convert the AC voltage in an unregulated DC voltage source.  The MOS devices in the rectifier are directly exposed to large voltage variations which depend on the proximity of the external reader. Two design considerations have been applied in order to protect these devices. First, 3.3 V transistors, which have a thicker oxide and higher breakdown voltage than their 1.8~V counterparts, have been used. Next, the PMOS diode chain composed of M5, M6, and M7 is used in a similar way to a Zener diode. The chain starts conducting at around 2~V and maintains the maximum unregulated voltage at around 3~V, while dissipating the excess of induced power. An additional consideration is that the pads connected to the coil terminals are by necessity disconnected from the ESD protection circuits. In this case, the PMOS diode chain provides an ESD protection mechanism to the devices in the rectifier since it also sinks currents  produced during ESD events. Finally, the switch M8 is used to transmit data to the reader by performing load modulation on the LC tank.

The unregulated voltage is stored on a 20~$\mu$F SM0603 external capacitor. This capacitor acts as an energy reservoir and it is necessary since during some instants of the serial data transmission, the LC tank is short-circuited when M8 is activated. During these short intervals, this capacitor slowly discharges while sourcing the necessary power to the circuits. When M8 is deactivated, the rectifier charges again the capacitor.

A low drop-out regulator (LDO) and a  1.22 V bandgap reference generate a 1.8~V power supply for the entire chip. A constant GM stage powered from the unregulated voltage line is used to provide biasing currents to both the LDO and the bandgap reference circuit. These currents have rather large variations, so they are only used in these two circuits. The biasing currents for the different blocks in the chip are instead derived from a current reference generated directly from the bandgap stage.

\subsection{Serial Data Reception}
\begin{figure}[t!]
\centering
\includegraphics[width=7.5 cm]{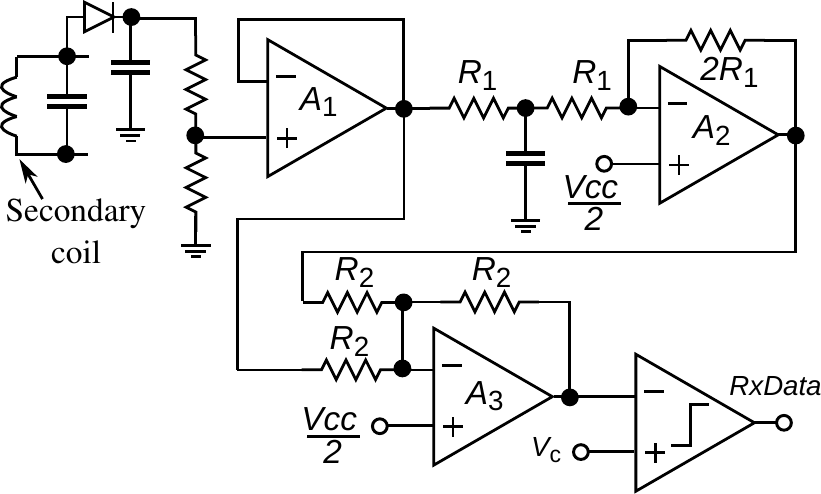}
\caption{ASK demodulator}
\label{FIG:DEMOD}
\end{figure}

Fig. \ref{FIG:DEMOD} shows the ASK demodulation circuits which is composed of a diode-based AM demodulator that is connected to one terminal of the secondary coil. The output of the demodulator is  followed by a buffer (A1).  The DC signal at the output of the demodulator depends on the proximity to the reader and could be in some cases as high as several volts. Accordingly, the input of the buffer is taken from a voltage divider so that the DC input level is within the 1.8~V range. The circuit configuration around A2 extracts the DC level of the demodulated signal and inverts its sign. Then, the voltage adder formed by A3 subtracts the buffered signal from its DC level, amplifies it, and centers the DC level at 0.9~V.  Finally, a comparator extracts the received data. The sensitivity of the demodulator is controlled by the comparison voltage $V_C$.

\subsection{Frequency Synthesis}
\begin{figure}[t!]
\centering
\includegraphics[width=7.5 cm]{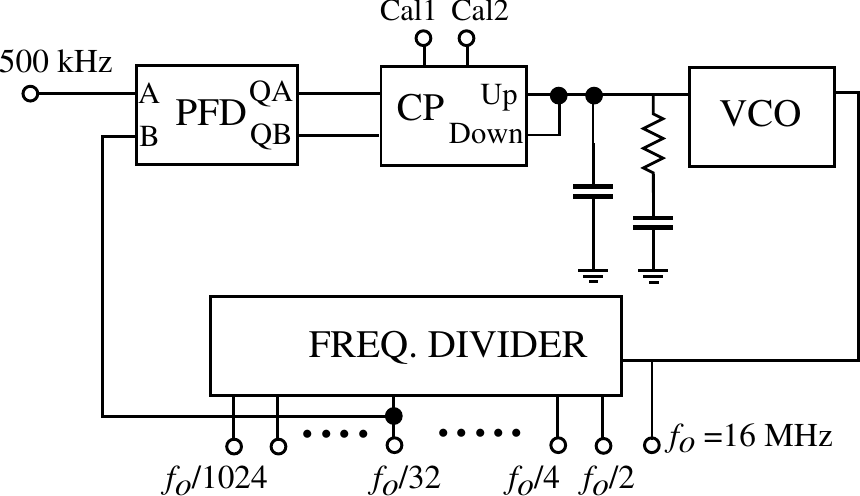}
\caption{32X Phase-locked loop }
\label{FIG:PLL}
\end{figure}

The frequencies needed by bio-impedance ASIC are generated by using the fully integrated 32X PLL frequency multiplier shown in Fig.~\ref{FIG:PLL}. A 500~kHz reference clock coming from the microcontroller produces a 16~MHz clock at the output of the PLL. The frequency divider is implemented by cascading 10 divide-by-two circuits made of D-type flip-flops. The buffered output of the VCO and the outputs of the dividers provide frequencies from 16~MHz down to 15.625~kHz. The SPI digital configuration block selects one of these frequencies by using a 16:1 multiplexer, as shown in Fig.~\ref{FIG:ARCHIT}. 

\begin{figure}[t!]
\centering
\includegraphics[width=8.5 cm]{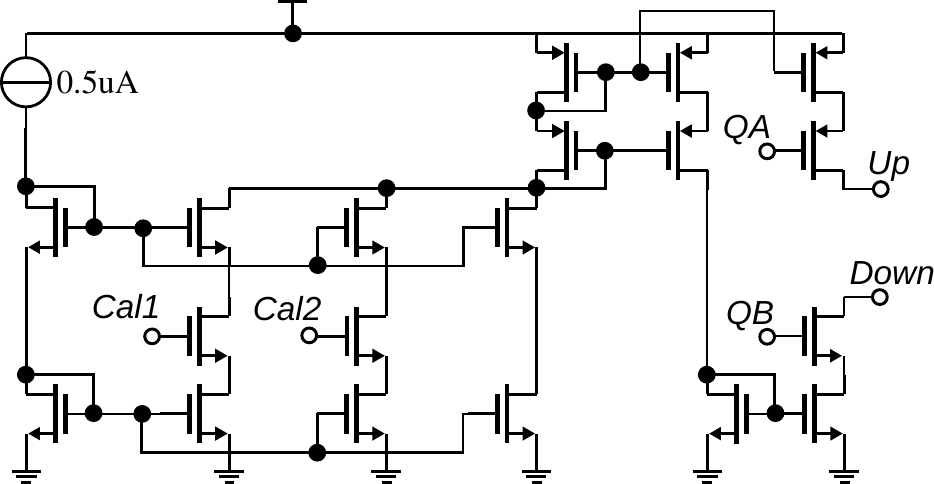}
\caption{Charge Pump }
\label{FIG:CP}
\end{figure}

The charge pump circuit is shown in Fig.~\ref{FIG:CP} and includes two calibration structures that allow to increase/decrease the charge pump's gain and therefore introduce a mechanism to control the loop dynamics of the PLL (lock speed, ringing, etc.). These calibration structures are directly controlled from the SPI digital configuration block.

\begin{figure}[t!]
\centering
\includegraphics[width=8.5 cm]{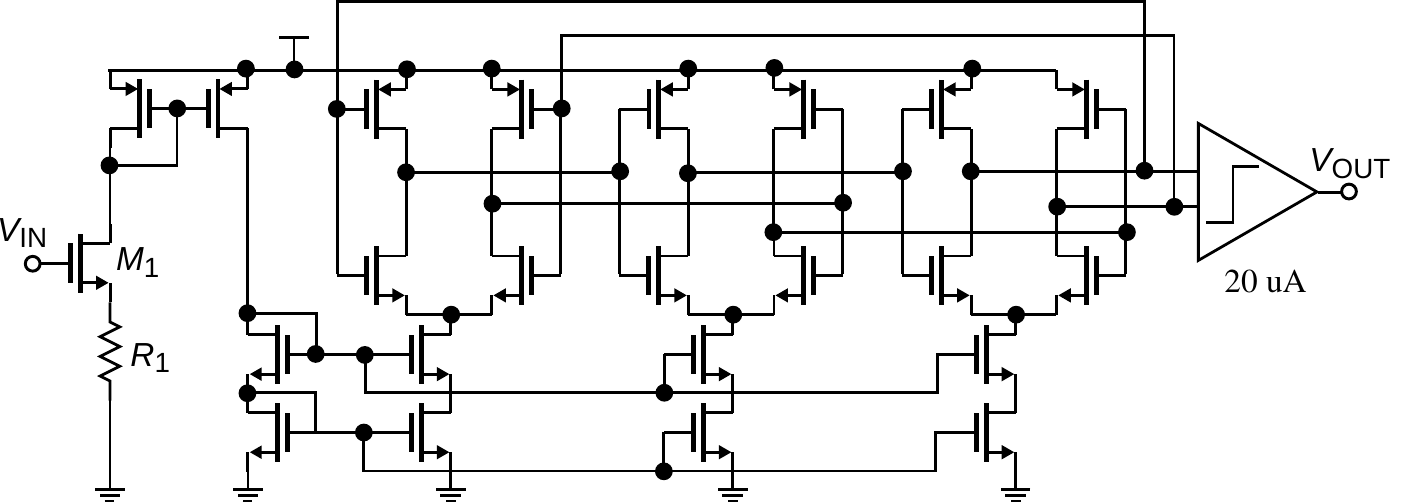}
\caption{VCO}
\label{FIG:VCO}
\end{figure}

As it is shown in Fig.~\ref{FIG:VCO}, the VCO consists of a 3-stage ring-oscillator in which the oscillation frequency is controlled by a transconductor stage formed by M1 and R1. A comparator takes the VCO signal and outputs a rail-to-rail square wave. The VCO's gain is around 22~MHz/V.

\subsection{Sinusoidal Signal and I/Q Generator}

\begin{figure}[t!]
\centering
\includegraphics[width=8.5 cm]{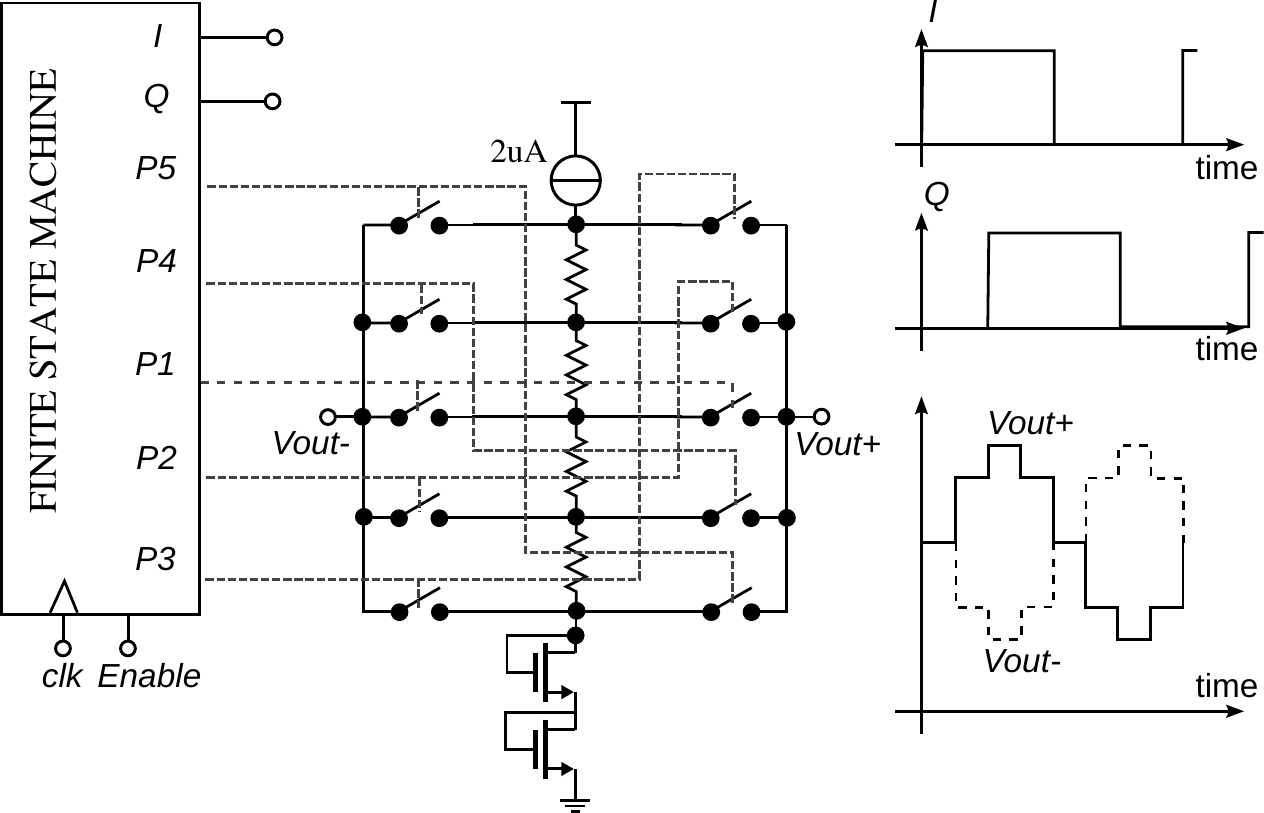}
\caption{Sinusoidal source, I/Q generation, and time diagrams}
\label{FIG:SOURCE}
\end{figure}

Fig.~\ref{FIG:SOURCE} shows the sinusoidal signal source and I/Q clock generators. The output of the 16:1 multiplexer, which consists of the selected frequency clock, is connected to the clk input of a finite state machine (FSM). This FSM outputs five control signals that are used to connect the output ports $V_{out+}$ and $V_{out-}$ to different nodes in a resistive voltage divider. The resistors are sized and matched so that an 8-step sinusoidal wave with an amplitude of 100~mV is available at the  differential  output $V_{out}$. The voltage divider is biased by 2~$\mu$A. Two NMOS diodes are used to set the common-mode voltage to a value close to 0.9~V. In addition, the FSM outputs in-phase and quadrature square waves. Due to the way in which all these signals are generated, the sinusoidal wave is always delayed $\pi/8$ rads (22.5$^{\circ}$) with respect to the I/Q signals. Hence, the measured impedance after I/Q demodulation  is rotated 22.5$^{\circ}$ in the complex plane. This is easily corrected later in software (either in the microcontroller or Android phone) by rotating the constellation with the same amount but in the opposite direction .

The Enable input of the FSM is used to activate/deactivate the sinusoidal signal generator. When it is deactivated, only the control signal P1 is set, which produces a differential $V_{out} = 0$. 

\begin{figure}[t!]
\centering
\includegraphics[width=9 cm]{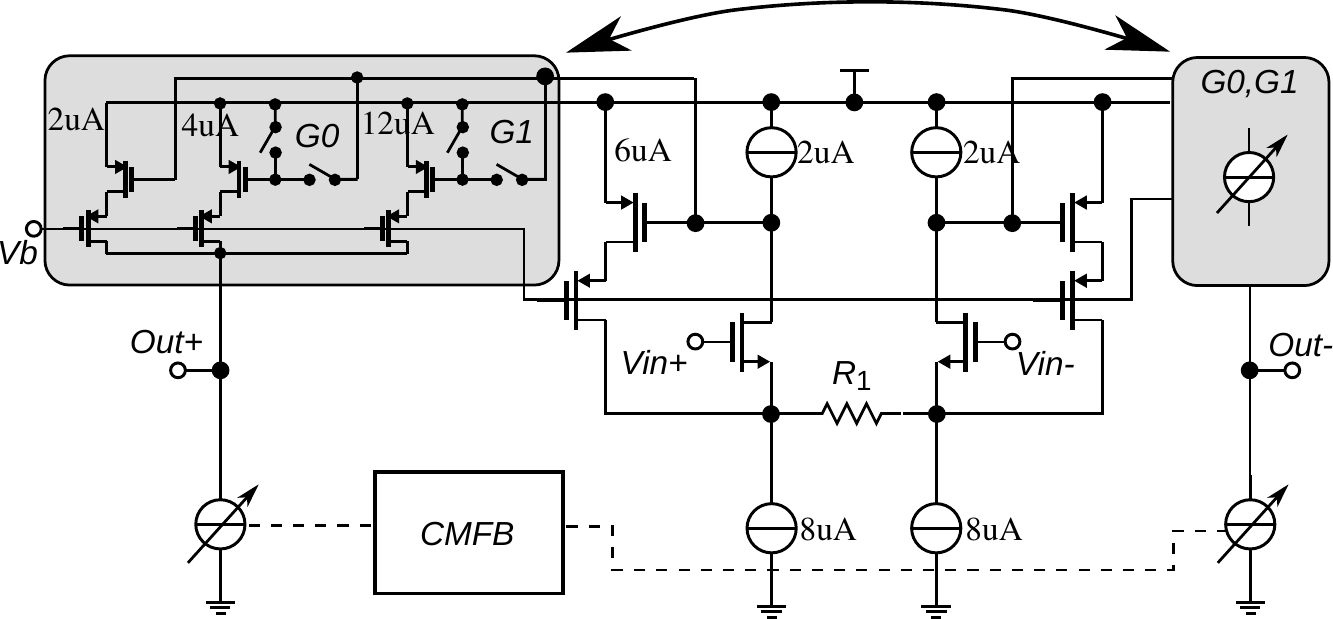}
\caption{Current signal generator}
\label{FIG:SGEN}
\end{figure}

The sinusoidal voltage is converted into current by using the variable transconductance circuit shown in Fig.~\ref{FIG:SGEN}. The $G0$ and $G1$ configuration bits allow the selection of three output current amplitude levels: 1.11~$\mu$A, 3.33~$\mu$A, and 10~$\mu$A. These differential currents are connected directly to the current injection electrodes. Both the voltage divider and the transconductance use the same type of poly resistor. Since the output current is proportional to the ratio of  the resistances, process drifts in the same direction are canceled.

\subsection{Voltage Amplification and  I/Q Demodulation}

\begin{figure*}[t!]
\centering
\includegraphics[width=14 cm]{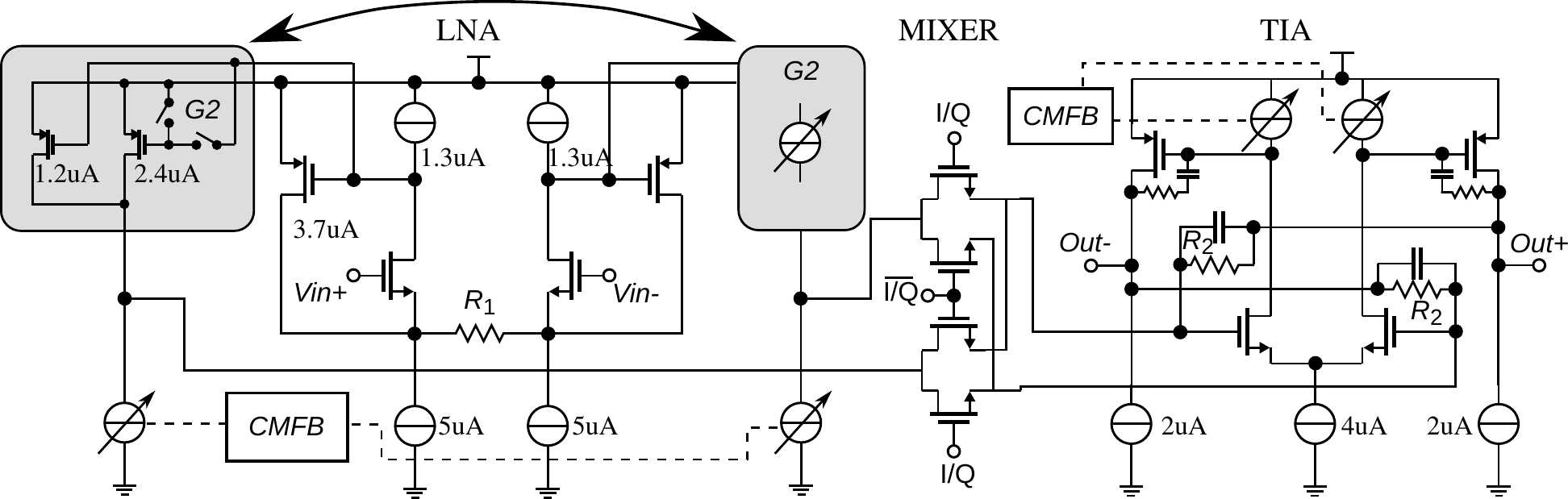}
\caption{Voltage amplification and I/Q demodulation}
\label{FIG:LNA}
\end{figure*}

Fig.~\ref{FIG:LNA} shows the voltage amplification and I/Q demodulation chain. The voltage at the sensing electrodes is converted into a differential  current by using a  low-noise variable transconductance amplifier. The configuration bit G2 controls the gain of the amplifier: when G2 is 1, GM = 20 $\mu$S, where when it is 0, GM = 6.67 $\mu$S. The LNA's output current is multiplied by the I or Q square wave by using a current commutating passive mixer. The output of the mixer is connected to a transimpedance amplifier (TIA) that both amplifies and partially removes high frequency components by performing first order low-pass filtering. 

Since the multiplication is done by using square wave mixing, a careful frequency planning is needed in order to avoid that the  harmonics of the square wave and the harmonics of the sinusoidal signal affect the final results.  In this work, the 8-step sinusoidal signal presents harmonics centered around frequencies which are multiples of 8 times the fundamental frequency (8X, 16X, 24X, etc). The main concern is the effect of the first harmonics which appear around 8 times the fundamental frequency since their level is only approximately 20 dB below the fundamental component. These harmonics are multiplied by the 7th and 9th harmonic of the square wave and their results appear superimposed on the desired result at DC with around 40 dB attenuation. Accordingly, they would introduce an error of approximately 1\% in the measurement. Nevertheless, most of the tissues behave as a low-pass filter in the $\beta$ region (10~kHz - 1~MHz), and therefore help to attenuate the high frequency harmonics of the injected signal before the mixing process takes place. As a result, the errors introduced due to mixing the 8-step sinusoidal signal and the I/Q clocks are below 1\%.

\subsection{Low-pass Filter and Buffer}

\begin{figure}[t!]
\centering
\includegraphics[width=8.5 cm]{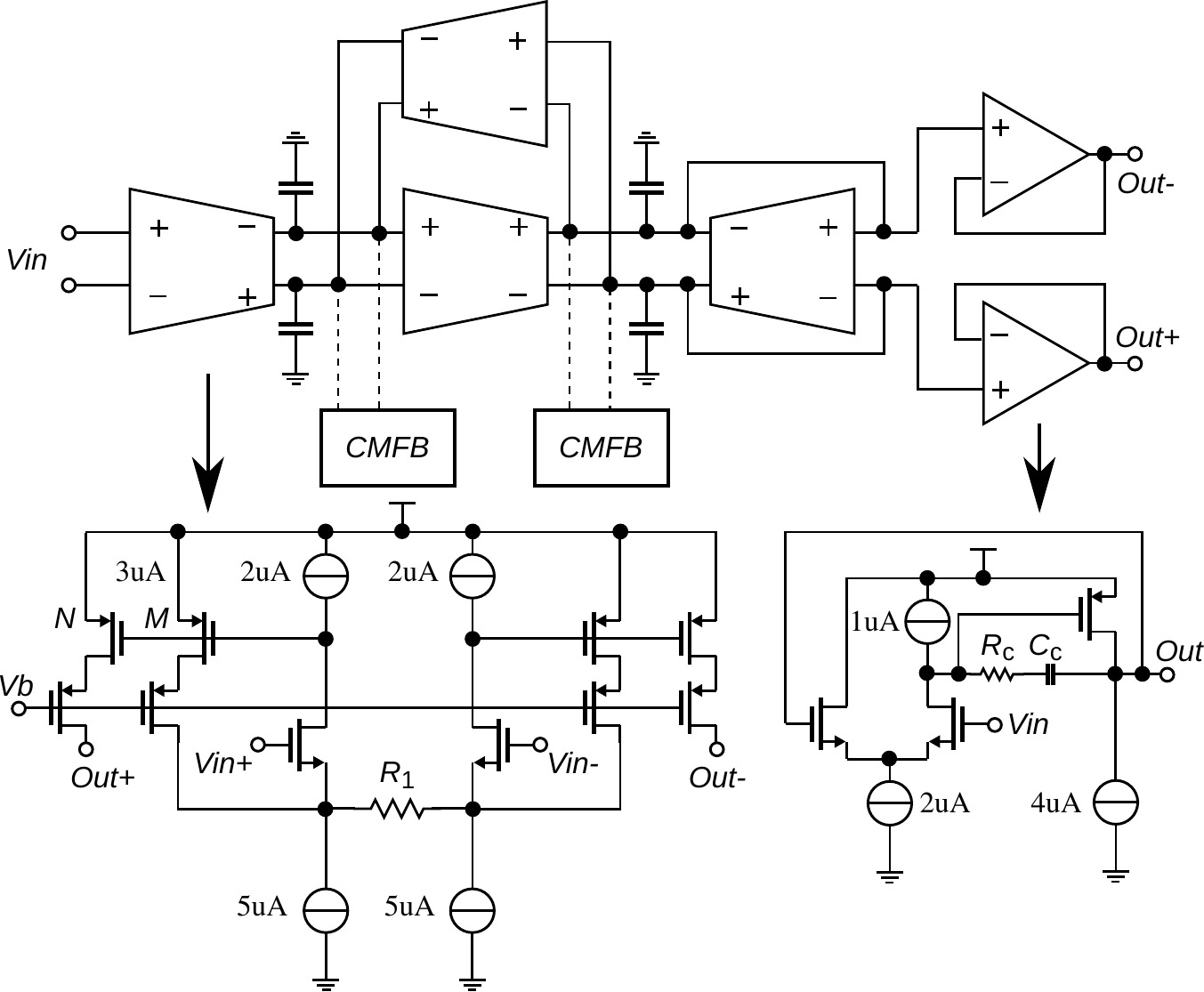}
\caption{Low-pass filter and buffer}
\label{FIG:FILTER}
\end{figure}

The results of the I/Q demodulation are purely DC signals and therefore high-frequency components coming from the mixing process need to be completely suppressed. A first order filtering is already provided by the TIA. The rest of the filtering is performed in the GM-C low-pass filter shown in Fig.~\ref{FIG:FILTER}. The filter consists of a second-order Chebyshev filter with a cut-off frequency of 50~Hz. Such low frequencies require external capacitors and extremely low GM values of around 0.1~$\mu$S. These transconductances   are achieved by using the ratio of the output current mirrors as an attenuator factor so that $GM = N/(MR_1), N<M$.

The filtered signal is connected to the output buffers which are designed to drive the output pads, PCB traces, and the sample-and-hold circuit of the ADC.

One serious concern in this kind of continuous-time implementation is the effect of $1/f$ noise on the final DC measurements. The following design considerations were applied here. First, the demodulated signal at the input of the filter is already amplified to at least a few hundred mV. This means that the input-referred integrated noise that can be tolerated to maintain the error level below 1\% is in the order of a couple of mVrms. Likewise, the quantization step of the 10-bit ADC in the microcontroller is 1.75~mV. Therefore, noise levels below this value are inoffensive, and design efforts to reduce $1/f$ noise below this value will only result in increased power consumption without any performance improvement. In this work, the filter and buffer were designed so that a simulated total output noise of around 1.3 mVrms is obtained after integrating the noise in the relevant bandwidth (1 Hz - 100 Hz). Another important observation is that the output of the I/Q demodulator is a DC signal whereas the $1/f$ noise is a random process with zero mean (no DC component). This gives the opportunity to mitigate the $1/f$ noise by just oversampling the output and applying averaging.

\subsection{Functional Timing Diagram}
\begin{figure}[h!]
\centering
\includegraphics[width=7cm]{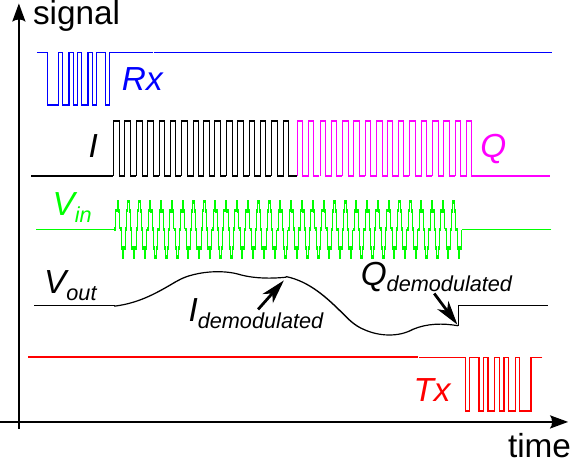}
\caption{Timing diagram }
\label{FIG:TIME_DIAG}
\end{figure}

{
Fig.~\ref{FIG:TIME_DIAG} shows the functional timing diagram of the implantable device including key waveforms from the ASIC. Once a serial command is received ($R_X$), the current signal generator and the $I$ reference are activated. A differential voltage drop on the tissue is sensed ($V_{in}$), amplified, and demodulated. The transient of the demodulated signal ($V_{out}$) settles in approximately 25~ms. The $DC$ signal corresponding to $I$ is sampled by the ADC several times, and averaged. Thereafter, the reference is switched to $Q$ and after the transient settles, the $DC$ signal corresponding to $Q$ is sampled by the ADC several times, and averaged. Finally, the current signal generator and the I/Q reference are disabled, and the information   transmitted to the reader ($T_X$).
}

\str{Table II shows a performance summary of the ASIC including the individual current consumption of each block.}

\begin{table}[!t]
\renewcommand{\arraystretch}{1.5}
\caption{ASIC Performance Summary}
\centering

\begin{tabular}{c c}

\hline\hline
Technology & 150 nm CMOS \\
Size & 1.22 mm $\times$ 1.22 mm \\
Power Supply & 1.8 V \\
Total Current & 165.5 $\mu$A\\

\hline
\multicolumn{2}{ c }{  Current Consumption Distribution} \\
\hline

Power Management + Comm. & 16.5 $\mu$A  \\
PLL &  22.9 $\mu$A\\
Signal Generator & 29.0 $\mu$A\\
LNA + Mixer & 17.4 $\mu$A\\
TIA & 9.2 $\mu$A\\
LPF & 58.5 $\mu$A\\
Buffers & 12.0 $\mu$A\\
\hline
\end{tabular}

\label{TABLE:SPEC}
\end{table}

\section{System Integration and Measurements}

\begin{figure}[t!]
\centering
\includegraphics[width=7.5 cm]{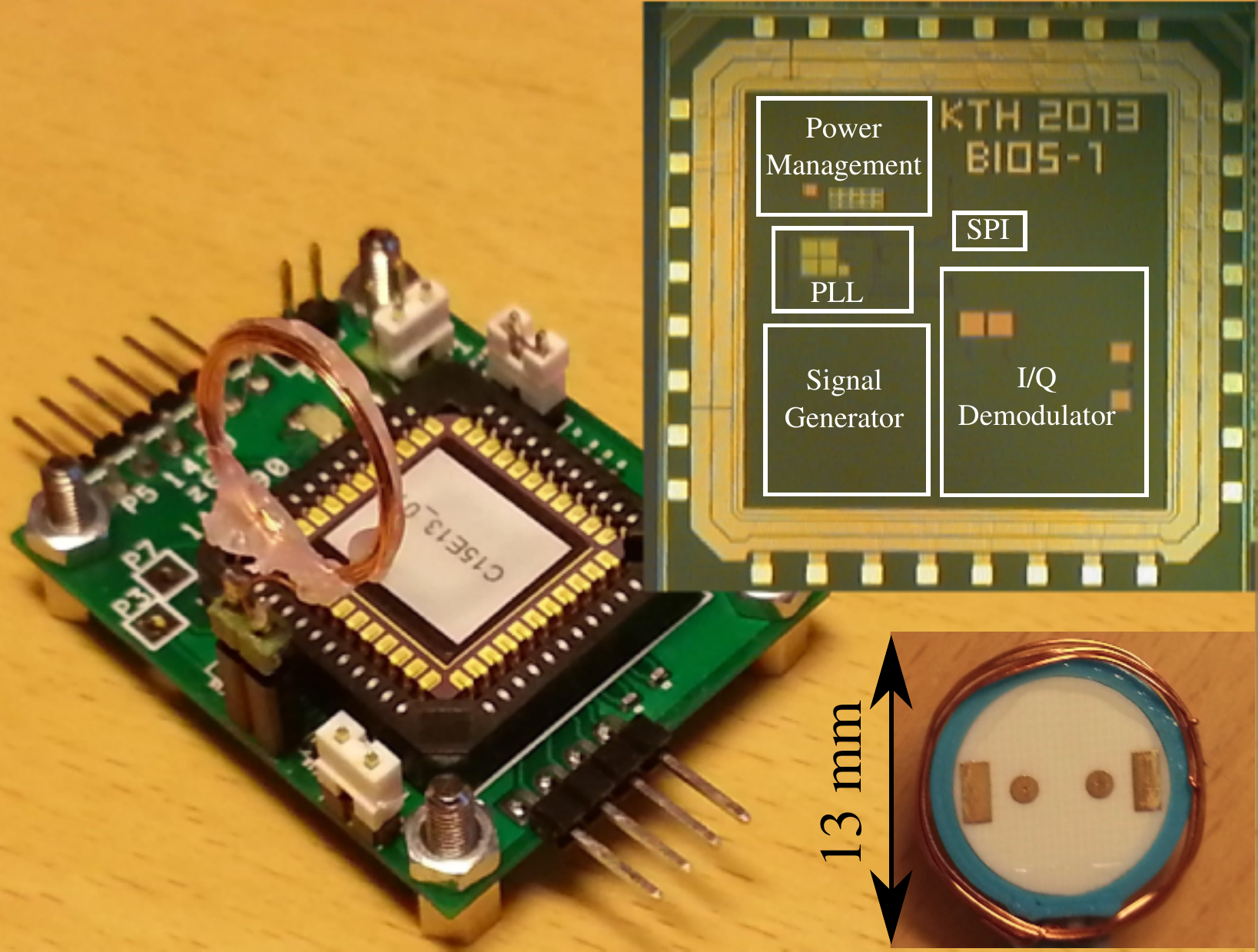}
\caption{Bio-impedance ASIC chip photo and test prototype}
\label{FIG:PROT}
\end{figure}

The bio-impedance ASIC was fabricated in a 150~nm 1.8~V CMOS process and bond-wired in a PLCC44 package for testing purposes. The circuit blocks occupy an active area of approximately 1.22~mm $\times$ 1.22 mm and consumes 165~$\mu$A. The PCB prototype for the implantable device including a micro-photograph of the ASIC is shown in Fig.~\ref{FIG:PROT}.
 { This PCB prototype is only used for testing and validation purposes, and includes testing points as well as programming circuitry for the microcontroller. A version ready for  \textit{in~vivo} testing contains all the SMD components including the ASIC mounted at the reverse of the electrodes. An inset of this version has been added at the bottom-right corner of Fig.~\ref{FIG:PROT} for comparison purposes. }

\subsection{Electrical Tests}

\begin{figure}[t!]
\centering
\includegraphics[width=7.5 cm]{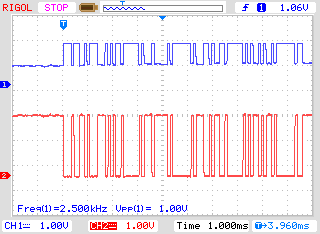}
\caption{Serial communication demodulated waveforms}
\label{FIG:MEAS2}
\end{figure}

Tests on powering and communication circuits show that the inductive coupling is able to power correctly the implantable device at distances up to 5~cm.  
{The power efficiency at this distance is around 0.3\%, which is enough to provide the minimum required power and communications to the device. At shorter distances, much higher efficiencies are achieved since the coupling coefficient increases. However, they are not relevant as the main objective of this system is to measure the bio-impedance of organs which can be located several cm under the skin.}
These tests have been performed while sending orders and receiving measurement data. This is important because power transfer is practically rendered ineffective during some instants of the communication. Fig. \ref{FIG:MEAS2} shows measured waveforms of the half-duplex communication at the reader when receiving data from the implantable prototype. The blue waveform is measured at the output of the envelope detector, whereas the red waveform is the 9.6~kbps serial data at the input of the USART of the Bluetooth module.

\begin{figure}[t!]
\centering
\includegraphics[width=7.5 cm]{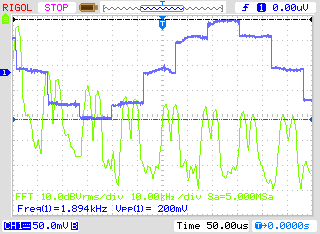}
\caption{Signal generator tests}
\label{FIG:MEAS1}
\end{figure}

Tests for validating the functionality of the signal generator are shown in Fig.~\ref{FIG:MEAS1}. The measured waveform of the 8-step sinusoidal signal is taken at one of the voltage sensing inputs. The measured spectrum of the waveform shows that the 7th and 9th harmonics are the strongest ones and are located around 20~dB below the fundamental.

Finally, the functionality of the I/Q demodulator is validated by measuring the following metrics: offsets, noise, linearity, and frequency response. 

DC offsets appear due to mismatches in the differential circuits. The DC offsets in the signal generator and LNA are up-converted by the mixer and removed by the low-pass filter. The TIA only process input currents, and therefore its input referred offset voltage (and also its input referred $1/f$ noise) is highly attenuated. Accordingly, the dominant contributor to the offset is the continuous-time filter.   The DC offset is superimposed to the result of the I/Q demodulator, and therefore, it should be carefully measured and compensated. Typically, a calibration loop is used in order to mitigate the offset. In this work, DC output offsets are first measured and subsequently subtracted from I/Q measurements by software. The offset measurement requires that the signal generator is disabled ($V_{IN}$ = 0 V), while the I/Q clock is active. The measured differential offset voltages in the tested samples are below 50 mV.

\begin{figure}[t!]
\centering
\includegraphics[width=7.5 cm]{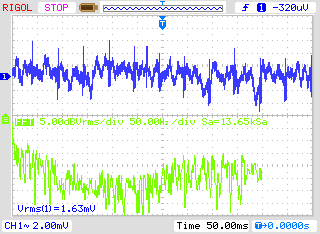}
\caption{Output noise waveform and spectrum}
\label{FIG:MEAS3}
\end{figure}

Fig.~\ref{FIG:MEAS3} shows the noise measured at one of the differential outputs which has a value of around 1.63~mVrms. The noise is dominated by $1/f$ noise coming from the filter. When taken differentially, it will slightly exceed the total output simulated noise; however, it is important to take into consideration that this is a single ended measurement, and therefore this measurement includes common-mode and power supply noise sources. As mentioned in previous sections, low frequency noise can be mitigated in DC measurements by taking multiple measurements and performing averaging. The averaging is performed in software where the number of taps is configurable (default number is 32). 

\begin{figure}[t!]
\centering
\includegraphics[width=7.5 cm]{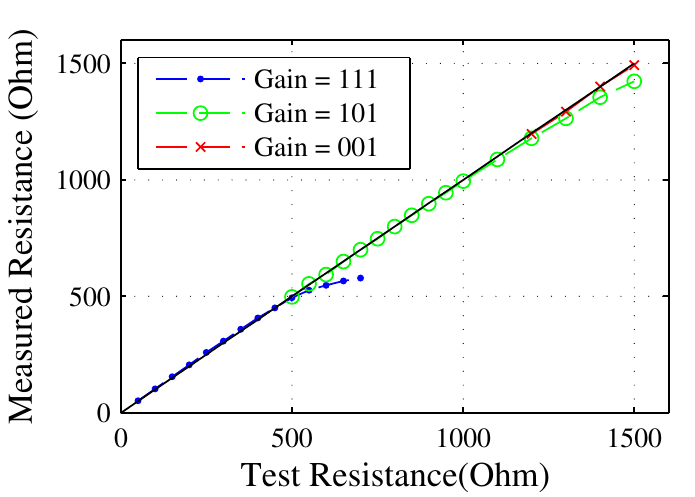}
\caption{Measured resistance values for different gains}
\label{FIG:MEAS4}
\end{figure}
 
The linearity of the bio-impedance sensor was evaluated by measuring resistors from 50 $\Omega$ to 2.5 k$\Omega$ (range of interest in this application). Such a large dynamic range requires  variable gains which are digitally controlled by the configuration  bits $G_0$, $G_1$, and $G_2$. Fig.~\ref{FIG:MEAS4} shows the linearity curves for different gain settings. The gain settings are defined as the digital word $G_0G_1G_2$. When all the gain bits are enabled (111), the response is linear up to around 400~$\Omega$. Then the system starts to exhibit compression and therefore it is necessary to switch to 101. This gain remains linear up to around 1.2~k$\Omega$ where it starts to compress. At this level it is necessary to switch the gain to 001, which guarantees very good linear response for levels up to around 3.6~k$\Omega$. In addition, higher impedances can be measured if $G_2$  is disabled. Although not designed for measuring high impedance tissues (skin or fat), disabling $G_2$ can extend the application of this ASIC to measure impedances up to around 11~k$\Omega$.

\begin{figure}[t!]
\centering
	\subfigure[]
	{
		\includegraphics[width=7.5 cm]{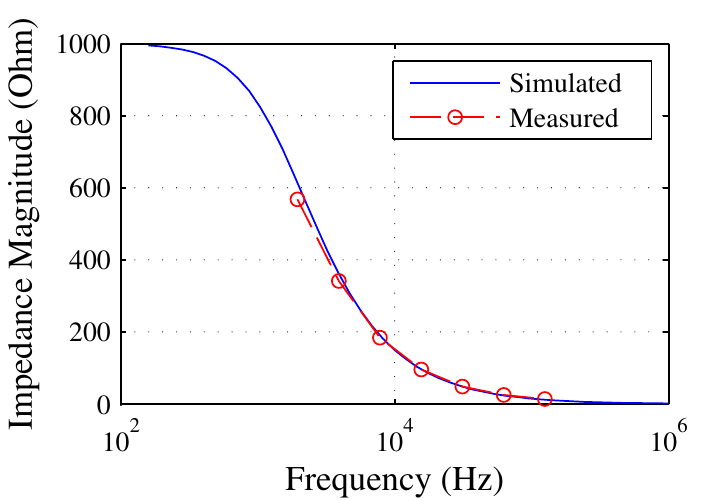}
	}
	\\
	\subfigure[]
	{
		\includegraphics[width=7.5 cm]{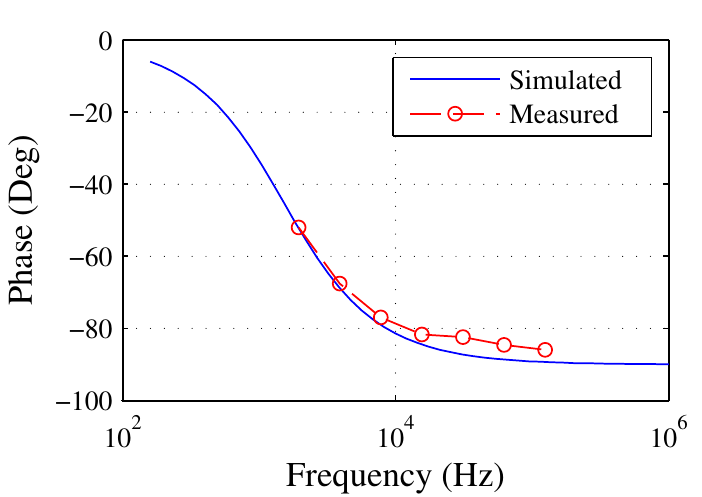}
	}

	\caption{Impedance magnitude (a) and phase (b) of 0.1 nF capacitor in paralell to 1 k$\Omega$ resistor}
\label{FIG:MEAS5}
\end{figure}

The frequency response was characterized by {assuming a ``Debye-type'' electrical model for the tissues. In this model, tissues can be represented by the parallel combination of a resistor and a capacitor. Different combinations of resistors and capacitors in the relevant range were measured. The values of resistors and capacitors were selected based on the results from Section II-A.} Fig.~\ref{FIG:MEAS5}(a) and Fig.~\ref{FIG:MEAS5}(b) show the impedance magnitude and phase for a 0.1 nF capacitor in parallel to a 1 k$\Omega$ resistor. At around 2 kHz, the impedance magnitude is close to 600 $\Omega$ whereas at 125~kHz the impedance drops to around 13~$\Omega$. Measurements at higher frequencies result in very small impedance values which are too close to the noise level and therefore they are disregarded. It is interesting to notice than even for a value as small as 13 $\Omega$, the phase error is still only a couple of degrees.

The previous measurements with discrete components provide valuable information of the system's performance. However, the electrical characteristics of biological tissues are defined by electrochemical properties and therefore an electrochemical characterization of the bio-impedance ASIC is more suitable.


\begin{figure}[t!]
\centering
	\subfigure[] {
		\includegraphics[width=7.5 cm]{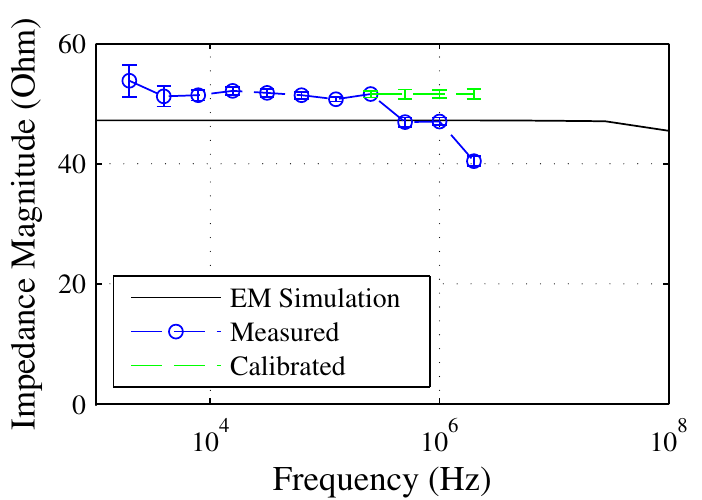}
	}
	\\
	\subfigure[] {
		\includegraphics[width=7.5 cm]{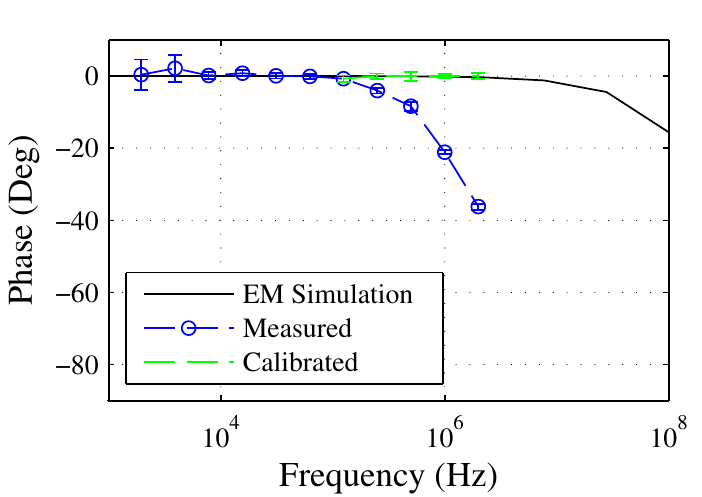}
	}
	\caption{Measured Impedance magnitude (a) and phase (b) of saline solution}
\label{FIG:MEAS7}
\end{figure}



\subsection{Electrochemical Tests}

Fig.~\ref{FIG:MEAS7}(a) and Fig.~\ref{FIG:MEAS7}(b) show measurements and simulations of impedance magnitude and phase for sterile physiological  solution (0.90 \% NaCl, $\sigma \approx$ 1.584 S/m, $\varepsilon_r \approx$ 78). The simulation was performed by extracting $Z_{21}$ parameters by using the EM simulation setup described in Section II. Ten measurements were taken for each frequency. The simulated impedance has  a magnitude of approximately 47 $\Omega$ at low frequencies and starts to drop at around 30~MHz. The measured impedance has a magnitude of approximately 52~$\Omega$ at low frequencies and starts to drop at around 250~kHz. It can be seen that the EM simulation follows very well the measurement at low frequencies with only around 5~$\Omega$  difference. This is a very encouraging result considering the rough approximations done during the EM simulation. On the other hand, the measurements start diverging from the simulation above 250~kHz. The differences are attributed to the first pole in the LNA which is located just above 2~MHz and that is low-pass filtering the signal. The filtering effect is reflected both in magnitude reduction and phase shift which are crucial parameters in the impedance measurements. Calibration mechanisms implemented in the analog domain have been proposed in order to tackle this problem \cite{Langlois2014}. In this work, since the offending pole is inside the amplification chain and totally isolated from the electrodes, it is possible to apply a correction mechanism based on equalization. The proposed calibration process consists on measuring a known impedance, i.e. 100 $\Omega$, and finding the calibration coefficients that are needed to correct each frequency component (both magnitude and phase). The calibration needs to be done only once, and when these coefficients are known, they are saved and can be reused to perform the correction in software (either in the microcontroller or in the Android phone). The green lines show the measurements that are calibrated using this method.

Finally, it can be seen that the standard error of the measurements is below 1 $\Omega_{rms}$ for frequencies above 4~kHz, and increases to 2.6 $\Omega_{rms}$ at 2~kHz. This increase is attributed to larger amounts of $1/f$ noise entering in the system. As discussed before, increasing the number of taps in the average filter reduces this noise to the quantization level at the expense of larger observation time.

\subsection{Ex Vivo Tests}

\begin{figure}[t!]
\centering
\includegraphics[width=8 cm]{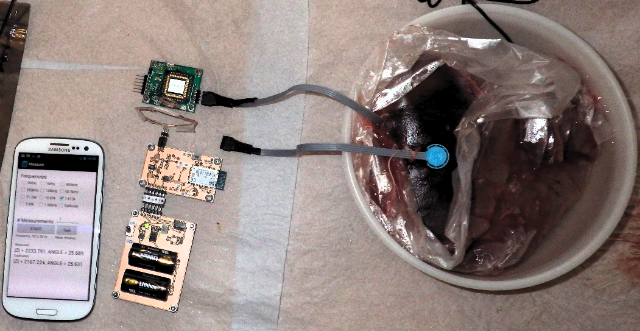}
\caption{Ex vivo test on sheep's liver and kidney}
\label{FIG:MEAS9}
\end{figure}

\begin{figure}[t!]
\centering
	\subfigure[] 
	{
		\includegraphics[width=8 cm]{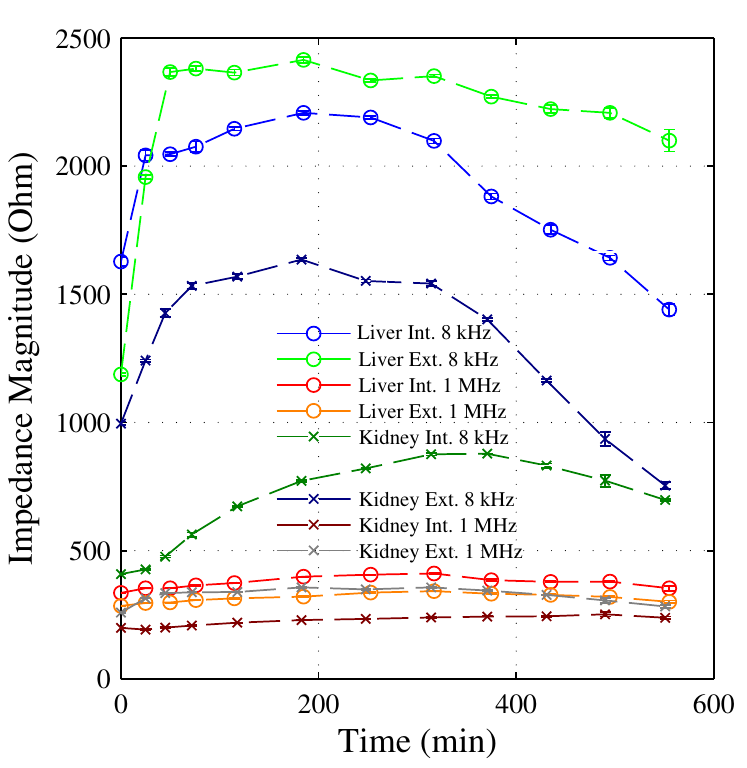}
	}
	\\
	\subfigure[]
	{
		\includegraphics[width=8 cm]{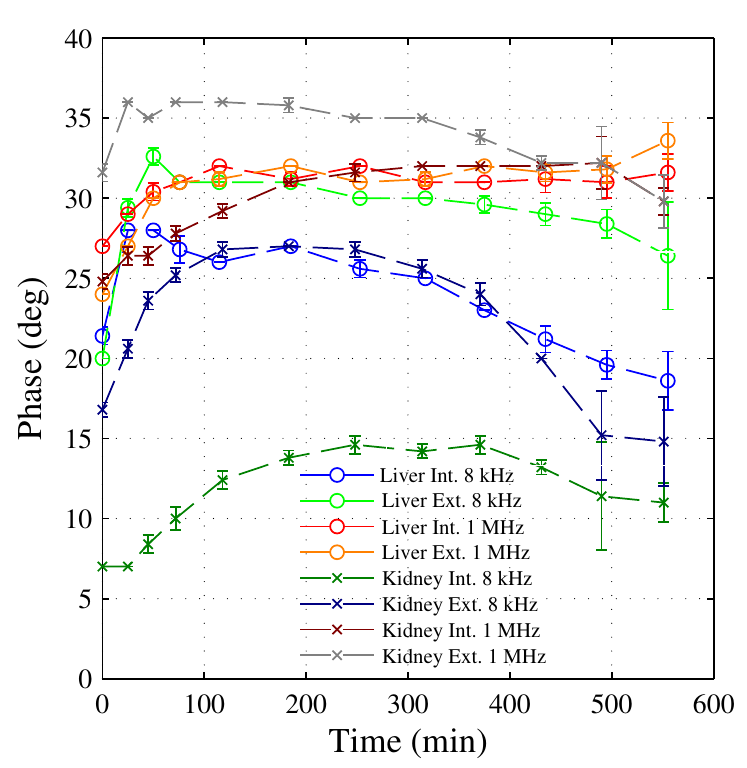}
	}
	\caption{Measured Impedance (a) and phase (b) of sheep's liver and kidney}
	\label{FIG:MEAS10}
\end{figure}

The final validation of the ASIC consists of performing \textit{ex~vivo} impedance measurements on sheep's liver and kidney at 8~kHz and 1~MHz (1 point in the lower half of the $\beta$ dispertion and 1 point in the upper end of the $\beta$ dispersion range of frequencies \cite{Grimnes2015}). \str{Fig.~\ref{FIG:MEAS9} shows the complete measurement setup: the Android mobile phone where the user application is installed, the external reader prototype running on batteries, the implantable device prototype, and the gold electrodes. }
 The measurement procedure was as follows. The measurements started 25 minutes after circulation stopped (Time zero in Fig.~\ref{FIG:MEAS10}(a) and Fig.~\ref{FIG:MEAS10}(b)), and lasted for several hours. A gold electrode probe was introduced in an incision done in each organ. In addition, another probe was fixed on the surface of the organs. The organs were deposited in plastic bags which were introduced in bowls filled with water. The water's temperature was constantly monitored and kept at around 37$^{\circ}$C.  Fig.~\ref{FIG:MEAS10}(a) and Fig.~\ref{FIG:MEAS10}(b) show the measured impedance's magnitude and phase respectively for the probes introduced in the incisions (Int.) and the ones attached externally (Ext.). It is observed that the magnitude at low frequencies increases a few hundreds of $\Omega$, remains relatively constant for some time,  and then decreases in some cases below its initial value. On the other hand, the magnitude at 1~MHz remains relatively constant  with values of a few hundreds of $\Omega$. The measured phase at 8~kHz follows the pattern of the measured impedance at the same frequency: first it increases, peaks for some time, and then it decreases. This pattern can be partially explained by noticing that for a very simple parallel RC model, an increase in R shifts the cut-off frequency to lower frequencies while increasing phase shift at higher frequencies. The behavior of the measured bio-impedances is in agreement with previous observations of ischemia in organs, where two factors inherent in R would be attributed to closing of gap junctions within a few hours after stop of circulation, followed later by rupture/lysis of cell membranes. A full decay of cell membranes would take another 10 hours or so, depending on temperature, and result in a lower impedance at the lower frequency than observed from the very beginning \cite{Gersing1998}. 

The \textit{ex~vivo} measurements confirm that the proposed ASIC accomplishes its target specifications, and therefore it can be successfully used to determine the bio-impedance of a variety of tissues in medical applications. The magnitude of the measured bio-impedances also confirm that the initial specifications, set for the minimum and maximum impedances, are at the correct levels. Furthermore, the measurements show that very accurate and stable measurements with a {resolution} of 1 $\Omega$ are possible even in \textit{ex~vivo} conditions.

\section{Conclusion}

This paper presented a 2-kHz to 2-MHz bio-impedance sensor ASIC that targets implantable biomedical applications. The ASIC is designed in 150~nm CMOS technology and consumes 165~$\mu$A at 1.8~V when powered by an external reader. The proposed ASIC has been validated by performing electrical, electrochemical, and \textit{ex~vivo} measurements.  All measurement results show that the proposed solution achieves around 1 $\Omega$ {resolution} when sensing a 100 $\Omega$ impedance (1\% error). In real medical applications, the tissues present larger impedance values; therefore, making possible better sensitivity levels. The measurement results show that this ASIC is able to successfully meet the bio-impedance sensing requirements while at the same time allowing a miniature size, battery-less implantable solution.

\section*{Acknowledgment}

The authors would like to thank Andreas Svensson, Martin Gustafsson (Maxim), Allan Olson (Exploric), H{\aa}kan Bengtsson, Adam Malmcrona, Jin Rizwand (Ericsson), and Jonas Fritzin (Intel Mobile Comm.) for the valuable comments during the design reviews, and Janko Katic, Tingsu Chen, and Tao Sha for help during the layout. Special thanks  go to the staff of Sk{\"a}rg{\aa}rdsslakteriet, V{\"a}rmd{\"o}, for allowing us to do \textit{ex~vivo} measurements in their premises.

\ifCLASSOPTIONcaptionsoff
  \newpage
\fi



\bibliographystyle{IEEEtran}
\bibliography{IMPLANTABLE_SENSORS}
%



%


\begin{IEEEbiography}[{\includegraphics[width=1in,height=1.25in,clip,keepaspectratio]{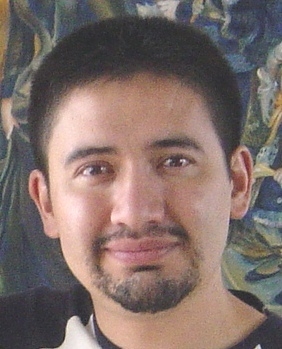}}]%
{Saul Rodriguez}
(M'06) received the B.Sc. degree in Electrical Engineering from the Army Polytechnic School (ESPE), Quito, Ecuador in 2001, the M.Sc. degree in System-on-Chip Design in 2005 and the Ph.D. degree in Electronic and Computer Systems in 2009 from the Royal Institute of Technology (KTH), Stockholm, Sweden. His research area covers from RF CMOS circuit design for wideband front-ends, ultra-low power circuits for medical applications, and graphene-based RF and AMS circuits.
\end{IEEEbiography}

\begin{IEEEbiography}[{\includegraphics[width=1in,height=1.25in,clip,keepaspectratio]{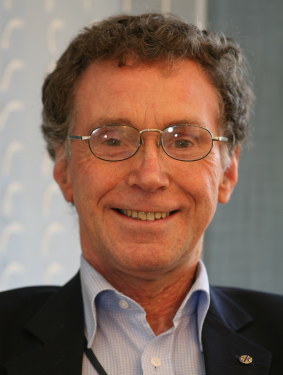}}]%
{Stig Ollmar}
(S'72-M'80,-SM'04) received his Ph.D. from the University of Stockholm in 1984 and then migrated stepwise to the Karolinska Institute where he has been Associate Professor since 1997. He is the founder of two medtec companies, one of which is now on the market with a diagnostic support tool $<$www.scibase.se$>$. Dr. Ollmar is half retired and enjoys singing in a choir, dancing, and playing electric bass. Interested in the history and philosophy of science, in particular the borderline between science and religion. Dr. Ollmar is a Senior Member of the IEEE.
\end{IEEEbiography}

\begin{IEEEbiography}[{\includegraphics[width=1in,height=1.25in,clip,keepaspectratio]{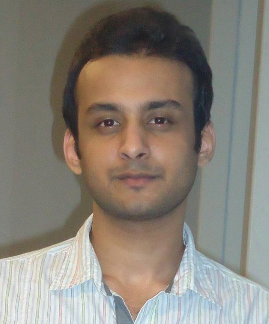}}]%
{Muhammad Waqar}
(S'15) received his M.Sc. degree in system on chip design from the KTH Royal Institute of Technology, Stockholm, Sweden, in 2012 where he worked in the design of implantable bio-medical circuits. He is currently working toward his Ph.D. degree in department of Integrated Devices and Circuits at ICT School in KTH. His current research interests include AC characterization and RF design of high temperature bipolar integrated devices and circuits.\end{IEEEbiography}

\begin{IEEEbiography}[{\includegraphics[width=1in,height=1.25in,clip,keepaspectratio]{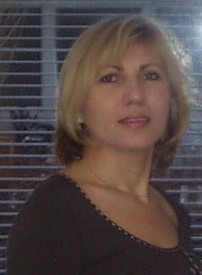}}]%
{Ana Rusu}
(M'92) received degrees of MSc (1983) in Electronics and Telecommunications and PhD (1998) in Electronics. Since September 2001, she has been with KTH Royal Institute of Technology, Stockholm, Sweden, where she is Professor in Electronics Circuits for Integrated Systems. Her research interests spans from low/ultra-low power high performance CMOS circuits and systems for a wide range of applications to circuits using emerging technologies, such as graphene and SiC. 
\end{IEEEbiography}






\vfill


\end{document}